\documentstyle[preprint,eqsecnum,aps]{revtex}
\begin{document}
\draft
\bibliographystyle{unsrt}
\preprint{USC(NT)-95-4}
\title{The $\pi NN$ Form Factor From QCD Sum Rules}
\author{T. Meissner}
\address{Department of Physics and Astronomy,\\
University of South Carolina,
Columbia, SC 29208, USA \\}
\date{\today}
\maketitle
\begin{abstract}
QCD sum rules are used to calculate the $q^2$ dependence of the $\pi NN$
coupling $g_{\pi NN} (q^2)$ in the spacelike region
$0.5 \ {\mbox{GeV}}^2 \lesssim q^2 \lesssim 1.5\ {\mbox{GeV}}^2$.
We study the Borel sum rule for the three point function of
one pseudoscalar and two nucleon currents up to order four
in the operator product expansion.
The Borel transform is performed with respect to the nucleon momenta,
whereas the momentum $q^2$ of the pseudoscalar
vertex is kept fixed at spacelike values.
The results can be well fitted using a monopole form
with a cutoff mass of about $\Lambda_\pi \approx 800 \ {\mbox{MeV}}$.
\end{abstract}
\pacs{13.75.Gx, 12.38.Lg, 13.75.Cs, 24.85.+p}

\newpage

\section{Introduction}
\label{sec1}

The pion nucleon form factor $g_{\pi NN} (q^2)$ plays a very important,
but not less controversial role in the  framework of $\pi N$ and $N N$
dynamics.

In general, meson baryon form factors are used in one boson exchange potentials
(OBEP) for the $N N$ force in order to account for the microscopic,
i.e. quark and gluon structure of the mesons and baryons at a given vertex,
and therefore provide a natural cutoff description for the interaction
potential at short distances.
Most of the realistic OBEP fit the data with a monopole form for the $\pi NN$
form factor
\begin{equation}
f_{\pi NN} (q^2) = { {g_{\pi NN}(q^2)} \over {g_{\pi NN} ({m_\pi}^2)}}
= { {{\Lambda_\pi}^2 - {m_\pi}^2 } \over {{\Lambda_\pi}^2 - q^2}}
\label{eq1}
\end{equation}
with a large monopole mass $\Lambda_\pi \gtrsim 1.3 {\mbox{GeV}}$
\cite{MHE}.
On the other hand almost all other hadronic theories advocate a much softer
form factor with $\Lambda_\pi \approx 500 - 950 {\mbox{MeV}}$:
Chiral soliton and quark models for the nucleon
\cite{Coh,KM,The,Bra,Alb,MG},
the Goldberger-Treiman discrepancy between $q^2 = 0$ and $q^2 = {m_\pi}^2$
\cite{GSS},
general considerations on the structure of the $\pi NN$
vertex \cite{ST},
charge exchange reactions \cite{EL},
threshold $\pi$ production \cite{Lee},
deep inelastic lepton nucleon scattering (Sullivan process) \cite{Tho,FMS,HSB}.
More recent studies within the OBEP, which either include the $\pi^\prime$
resonance \cite{HT}, use a different form of the scalar exchange potential
\cite{Dei} or consider a correlated $\rho \pi$exchange
\cite{Jan} also indicate a
softer cutoff mass ($\Lambda_\pi \approx 700 - 800 {\mbox{MeV}}$).

For these reasons it seems highly desirable to perform a
calculation of $g_{\pi NN} (q^2)$, which does not suffer from the ambiguities
of the model and parameterization schemes mentioned above and is connected
to QCD as closely as possible.

In this context a quenched lattice QCD calculation has recently been
carried out rendering a monopole cutoff mass of
$\Lambda_\pi = 750 {\mbox{MeV}}$ \cite{Liu}.

It is the aim of our work to study this phenomenon within the framework of QCD
sum rules \cite{SVZ},
which have turned out to be a very successful method for
calculating hadronic properties at
intermediate energies (for reviews c.f.\cite{RRY1,Nar})
without employing the computer time consuming lattice gauge calculations.

In order to calculate the pion nucleon coupling constant
$g_{\pi NN} = g_{\pi NN} (m_\pi ^2)$ itself,
i.e. for on shell $\pi$, one can consider either:
(i) the vacuum three point correlator of two nucleon and one pseudoscalar
meson interpolating fields, which are saturated with resonances in the nucleon
$N$ and pion $\pi$ channels on the phenomenological side
\cite{RRY1,RRY2,Rei,NP}
or (ii) the 2 point function of 2 nucleon interpolating fields sandwiched
between the vacuum and one $\pi$ state and saturating only with $N$
resonances \cite{RRY1,Rei,SH}.
With both methods it was possible to obtain rather reasonable results for
$ g_{\pi NN}$, although it should be stated that the uncertainties in
the Borel analysis are relatively high, especially in method (ii),
even if higher order power and $\alpha_s$ corrections are taken into
account \cite{SH}.

If one is interested in the momentum dependence of $g_{\pi NN} (q^2)$
at intermediate $q^2$ and therefore with off shell $\pi$, only method
(i), which has been successfully used for the calculation of meson
formfactors \cite{IS1,NR,IS2}, can be applied.

Our paper is organized as follows:
In section \ref{sec2} we introduce the 3 point function for the $NN\pi$
vertex and saturate with nucleon intermediate states.
In section \ref{sec3} the operator product expansion (OPE) is carried out.
In section \ref{sec4} we perform the Borel analysis. The results are
discussed
and summarized in section \ref{sec5}.

\section{The Three Point Function for the \protect{$\pi NN$} Vertex}
\label{sec2}

Our starting point is the 3 point function (Fig.\ref{fig1})

\begin{equation}
A(p_1,p_2,q) = \int d^4 x_1 d^4 x_2 e^{ip_1 x_1} e^{- ip_2 x_2}
\left \langle 0 \vert {\cal T}
\eta (x_1) J_5 ^0 (0) {\bar{\eta}} (x_2) \vert 0 \right \rangle
\label{eq2}
\end{equation}

of a pseudoscalar, charge neutral current
\begin{equation}
J_5 ^0 (x) =
{\bar{q}} (x) i \gamma_5 \tau^0 q ( x)
\label{eq3}
\end{equation}
and two Ioffe nucleon interpolating fields \cite{Iof}
\begin{equation}
\eta (x) = \epsilon_{abc} \left [ \left ( u^a (x) {\cal C} \gamma_\mu
u^b (x) \right ) \gamma_5 \gamma^\mu d^c (x) \right ]
\label{eq4}
\end{equation}
Hereby $q = {u \choose d}$ denote the spinors for the quarks
with mass $m_0 = m_u =m_d$,
$a,b,c$ the color indices
and ${\cal C} = i \gamma_2 \gamma_0$ the charge conjugation matrix.
The three momenta at the vertex are related by $ q = p_1 - p_2$.

Due to restrictions from Lorentz, parity and charge conjugation invariance
the 3 point function $A(p_1,p_2,q)$ has the general form
\begin{eqnarray}
A(p_1,p_2,q) =
& {F_1} & (p_1 ^2 , p_2 ^2 ,q^2) \gamma_5 +
F_2 (p_1 ^2 , p_2 ^2 ,q^2) \not{\! q} \gamma_5 + \nonumber\\
& {F_3} & (p_1 ^2 , p_2 ^2 ,q^2) \not{\! P} \gamma_5 +
F_4 (p_1 ^2 , p_2 ^2 ,q^2) \sigma_{\mu\nu}  \gamma_5 p_1 ^\mu p_2 ^\nu
\label{eqf}
\end{eqnarray}
where $q = p_1 -p_2$, $P = {{p_1 + p_2}\over 2}$.
The functions $F_1$,$F_2$,$F_4$ are symmetric and $F_3$ is antisymmetric
under the interchange $p_1 \leftrightarrow p_2$.

The matrix element of the pseudoscalar current $J_5 ^0$ between on shell
nucleon states defines the pseudoscalar nucleon formfactor
\begin{equation}
\left \langle N(p_1) \vert J_5 ^0 (0) \vert N(p_2) \right \rangle
 = g_P (q^2)
{\bar{u}}_N (p_2) i \gamma_5 u_N (p_1)
\label{eq5}
\end{equation}
where $u_N (p)$ denotes a free nucleon spinor in momentum space.

The pion nucleon coupling constant $g_{\pi NN}$ is defined by the
$\pi N$ interaction:
\begin{equation}
{\cal L}_{\pi N} = i g_{\pi NN} {\bar{N}} i \gamma_5 \bbox{\tau}\bbox{\pi} N
\label{eq5a}
\end{equation}
with the form factor $g_{\pi NN} (q^2)$
For on shell nucleons it is equivalent to use
the coupling of eq.(\ref{eq5a}) or the chirally invariant
vector coupling
\begin{equation}
{\cal L}_{int} = \frac{g_{\pi NN}}{2 M_N}
\partial_\mu \bbox{\pi} \left ( {\bar{N}} \gamma^\mu \gamma_5
\bbox{\tau} N
\right )
\label{eq5b}
\end{equation}

By means of the chiral Ward identity
\begin{equation}
\partial_\mu \bbox{ A}^\mu = m_0 {\bar q} (x) i \gamma_5 \bbox{\tau} q (x)
\label{eq6}
\end{equation}
for the axial current
$\bbox{ A}^\mu (x) = {\bar q} (x) \gamma^\mu \gamma_5
\frac{1}{2} \bbox{\tau} q(x)$
as well as the PCAC relation
\begin{equation}
\partial_\mu \bbox{ A}^\mu (x) = m_\pi ^2 f_\pi \bbox{\pi} (x)
\label{eq7}
\end{equation}
one can relate the pseudoscalar nucleon form factor $g_P (q^2)$
(eq.\ref{eq5}) to the pion nucleon form factor $g_{\pi NN} (q^2)$:
\begin{equation}
g_P (q^2) =
\left ( {{m_\pi ^2 f_\pi}\over {m_0}} \right )
{ {g_{\pi NN} (q^2)} \over { - q^2 + m_\pi ^2 } }
\label{eq8}
\end{equation}

Therefore saturating the eq.(\ref{eq2}) with nucleon states
$\vert N(p) \rangle$
renders, after continuing to Euclidean momenta
($q^2 \rightarrow - q^2$,
$p_1 ^2 \rightarrow - p_1 ^2$,
$p_2 ^2 \rightarrow - p_2 ^2$)
\begin{equation}
[A(p_1,p_2,q)]_N  =   i \lambda_N ^2
\left ( {{m_\pi ^2 f_\pi}\over {m_0}} \right ) \:
{ {g_{\pi NN} (q^2)} \over { q^2 + m_\pi ^2 } } \:
{{\left [
(\not{\! p_1} + M_N ) \gamma_5 (\not{\! p_2} + M_N )
\right ]}
\over{ (p_1 ^2 + M_N ^2) (p_2 ^2 + M_N ^2) }}
\label{eq9}
\end{equation}
where $\lambda_N$ is the overlap between the Ioffe current $\eta (x)$
and a nucleon state
\begin{equation}
\left \langle 0 \vert \eta (x) \vert N (p) \right \rangle = \lambda_N
e^{-ipx} u_N (p)
\label{eq10}
\end{equation}
Using $p_1 ^2 = p_2 ^2 = M_N ^2$
the $[ \dots ]$ term in eq.(\ref{eq9}) can be cast into the form
\begin{equation}
(\not{\! p_1} + M_N ) \gamma_5 (\not{\! p_2} + M_N )  =
(- p_1 \cdot p_2 + M_N^2 ) \gamma_5 + i \sigma_{\mu\nu} p_1 ^\mu p_2 ^\nu
\gamma_5 + M_N \not{\! q} \gamma_5
\label{eq11}
\end{equation}
The contribution of the first higher resonance $N^*$
with mass $M_N ^*$ to eq.(\ref{eq2})
has the form
\begin{equation}
[A(p_1,p_2,q)]_{N^*}   =   i \lambda_N \lambda_{N^*}
\left ( {{m_\pi ^2 f_\pi}\over {m_0}} \right ) \:
{ {g_{\pi N {N^*}} (q^2)} \over { q^2 + m_\pi ^2 } } \:
{{\left [
(\not{\! p_1} + M_N ) \gamma_5 (\not{\! p_2} + M_N ^* )
\right ]}
\over{ (p_1 ^2 + M_N ^2) (p_2 ^2 + {M_N ^*}^2) }}
\quad  +  \quad N \leftrightarrow N^*
\label{eq12}
\end{equation}

\section{Operator Product Expansion}
\label{sec3}

As already stated in eq.(\ref{eqf}) there are essentially 4 Lorentz structures
contained in $A(p_1,p_2,q)$.
In the following we work with massless quarks, i.e $m_0 =0$.
{}From a simple dimensional analysis of both sides of the sum rule
it is easy to see \cite{RRY1,RRY2,Rei} that in the
sum rule for the functions $F_1$ and $F_4$, which both contain
an even number of external momenta, only operators of even
dimension ($\openone$, $\langle G^2 \rangle$,
$ \langle \bar{q} \Gamma q \bar{q} \Gamma q \rangle$, etc.) contribute,
whereas the odd dimension operators enter with a mass factor $m_0$ and
therefore vanish for $m_0 \to 0$.
On the other hand for the functions $F_2$ and $F_3$ containing an odd number of
external momenta, only operators of odd dimension
($\langle \bar{q} q \rangle$, $\langle \bar{q} G \cdot \sigma q \rangle$)
contribute.
By multiplying both sides of the sum rule with $\not{\! q}$ and taking
the trace over the Dirac matrices, we can single out the odd dimensional
structure, i.e. the sum rule for $F_2$ and $F_3$.
Up to order 4 we only have to account for the diagrams in Fig.\ref{fig2}
containing the quark condensate
$\langle \bar{q} q \rangle = \langle {\bar u} u \rangle =
\langle \bar{d} d \rangle$.

The contribution from the diagrams in Fig.\ref{fig2}a renders:
\begin{equation}
\left [ A (p_1,p_2,q) \right ] _{2a} =
16 \, \gamma_5 \, \langle \bar{q} q \rangle \,
{ { \not{\! q} } \over {q^2} } \, {1\over 2} \,
\left [ p_1 ^2 I (p_1 ^2) + p_2 ^2 I(p_2 ^2) \right ]
\label{eq13}
\end{equation}
whereas for the diagrams of Fig.\ref{fig2}b one obtains after a
rather lengthy calculation:
\begin{eqnarray}
\left [ A (p_1,p_2,q) \right ] _{2b} \, = \,
 ( &-& 16 ) \, \gamma_5 \, \langle \bar{q} q \rangle \,
{ { \not{\! q} } \over 4 } \,
\left [
(p_1 ^2 + p_2 ^2) J(p_1 ^2, p_2 ^2, q^2) +
(p_1 ^2 - p_2 ^2) K_+ (p_1 ^2, p_2 ^2, q^2)
\right ] \, +
\nonumber\\
 ( &-& 16 ) \, \gamma_5 \, \langle \bar{q} q \rangle \,
{ { \not{\! P} } \over 2 } \,
\left [
(p_2 ^2 - p_1 ^2) J(p_1 ^2, p_2 ^2, q^2) +
(p_1 ^2 - p_2 ^2) K_- (p_1 ^2, p_2 ^2, q^2)
\right ] \, +
\nonumber\\
 ( &-& 16 ) \, \gamma_5 \, \langle \bar{q} q \rangle \,
{ { \not{\! q} } \over 4 } \,
\left [
{1\over 2} I(p_1 ^2) +
{1\over 2} I(p_2 ^2) +
2 I (q^2)
\right ] \, +
\nonumber\\
 ( &-& 16 ) \, \gamma_5 \, \langle \bar{q} q \rangle \,
{ { \not{\! P} } \over 2 } \,
\left [
{1\over 2} I(p_1 ^2) -
{1\over 2} I(p_2 ^2)
\right ]
\label{eq14}
\end{eqnarray}
The expressions $I$,$J$,$K_+$ and $K_-$ arise from loop integrations
and can be most conveniently represented as double parameter integrals
\begin{equation}
I(p^2) =
\int {{d^4 k} \over {(2 \pi)^2}}
{1\over{(p-k)^2 k^2}} = \left ( - {i\over{16 \pi ^2}} \right ) \ln \left (
 - {p^2 \over \mu^2} \right )
\label{eq15}
\end{equation}
\begin{eqnarray}
J(p_1 ^2, p_2 ^2, q^2) &=&
\int {{d^4 k} \over {(2 \pi)^2}}
{1\over{ k^2 (k - p_1)^2 (k - p_2)^2 }}
\nonumber\\
&=&
 \left ( - {i\over{16 \pi ^2}} \right )
(-)
\int_0^1 d \rho \int_0^1 d \lambda
{1 \over
{p_1 ^2 (1-\lambda) (1 - \rho)
+ p_2 ^2 (1-\lambda) \rho +
q^2 \lambda \rho (1 - \rho) }}
\label{eq16}
\end{eqnarray}
and
\begin{equation}
\int {{d^4 k} \over {(2 \pi)^2}}
{{\not{\! k}} \over{ k^2 (k - p_1)^2 (k - p_2)^2 }}
=
\left ( - {i\over{16 \pi ^2}} \right )
\left [
{{\not{\! q}}\over2}
K_+ (p_1 ^2, p_2 ^2, q^2)
+ \not{\! P}
K_- (p_1 ^2, p_2 ^2, q^2)
\right ]
\label{eq17}
\end{equation}
with
\begin{equation}
K_\pm (p_1 ^2 , p_2 ^2 ,q^2) =
\int_0^1 d \rho \int_0^1 d \lambda
{{ \lambda\rho \pm (\lambda -1) }
\over
{
p_1 ^2 \lambda \rho (1-\lambda)
+ p_2 ^2 (1-\lambda) (1 - \rho) +
q^2 (1- \lambda) \rho  }}
\label{eq18}
\end{equation}
In the UV divergent integral of eq.(\ref{eq15}) we have applied the standard
dimensional renormalization at the renormalization point $\mu$.

\section{Sum Rule and Borel Analysis}
\label{sec4}

For the OPE to be valid $q^2$ has to large, i.e. $q^2 \gg \Lambda_{QCD} ^2$.
In this case we can neglect the pion mass in the pole term in eq.(\ref{eq13}),
which is consistent with putting $m_0 =0$.

In refs.\cite{RRY1,RRY2,Rei} a sum rule for the pion nucleon coupling
$g_{\pi NN} = g_{\pi NN} (q^2 =0)$ has been obtained and analyzed
by identifying the residua of the $1\over q^2$ pole in eq.(\ref{eq9})
and in the OPE contribution from Fig.\ref{fig2}a (\ref{eq13}).
By introducing
\begin{equation}
\Delta g_{\pi NN} (q^2) = g_{\pi NN} (q^2) - g_{\pi NN} (0)
\label{eq19}
\end{equation}
and analogous expressions for the higher resonance contributions, we
therefore can write down the sum rule:
\begin{equation}
i \lambda_N ^2
\left ( {{m_\pi ^2 f_\pi}\over{m_0}} \right )
{ M_N \over
{
(p_1 ^2 + M_N ^2 )
(p_2 ^2 + M_N ^2 ) }}
\Delta g_{\pi NN} (q^2)
\quad
+ \quad \dots
\quad   = \quad
{1\over {4 q^2}}
{\rm Tr}_\gamma
\left [ \not{\! q} A (p_1 , p_2 , q)
\right ]_{2b}
\label{eq20}
\end{equation}

Following refs.\cite{IS1,NR,IS2} we apply the double Borel transform
with respect to the nucleon momenta $p_1 ^2$ and $p_2 ^2$
\begin{equation}
{\cal B}_{12} = {\cal B}_1 \cdot {\cal B}_2
\end{equation}
\begin{equation}
{\cal B}_i =
\lim_{
{
{\scriptscriptstyle
p_i ^2 \to \infty \, ,\,
n_i \to \infty} \atop
 {\scriptscriptstyle
M_i ^2 = \frac{p_i ^2}{n_i} \, {\rm fixed} } } } \,
{1 \over{(n_i -1) !}} \, (p_i ^2) ^{n_i} \,
\left (
- { \partial \over{\partial p_i ^2 }}
\right ) ^{n_i} \; , \quad i=1,2
\label{eq21}
\end{equation}
and put $M_1 ^2 = M_2 ^2 = M^2$.
The momentum $q^2$ of the pion is kept fixed at a
spacelike value $q^2 \gg \Lambda_{QCD} ^2$, so that both sides of the
sum rule depend on $M^2$ and $q^2$ now.

After applying the operator ${\cal B}_{12}$
on the OPE side the terms which depend only on either $p_1 ^2$ or
$p_2 ^2$ vanish.
The Borel transform of the other expressions can be performed by using
the relations
\begin{eqnarray}
&{\cal B}& \left \{ {1\over {p^2 + s}} \right \}
= {1\over M^2} e^{-{s\over M^2}}
\nonumber\\
&{\cal B}& \left \{ {p^2 \over {p^2 + s}} \right \}= { - {s\over M^2} }
e^{-{s\over M^2}}
\nonumber\\
&{\cal B}& \left \{ (p^2)^k e^{-\alpha p^2} \right \}
= ( - )^k {1\over M^2} \delta^{(k)} (\alpha - {\scriptstyle {1\over M^2} })
\label{eq22}
\end{eqnarray}
Taking the first resonance with the quantum numbers of the nucleon
$N^*$ explicitly into account in addition to
the ground state $N$ we obtain the Borel sum rule
\begin{eqnarray}
& {\lambda_N ^2} &
\left ( {{m_\pi ^2 f_\pi}\over{m_0}} \right ) \,
{M_N \over M^6} \,
e^{ -2 \frac{M_N ^2}{M^2}} \,
\left ( \Delta g_{\pi NN} (q^2) \right ) \, +
\nonumber\\
2  &{\lambda_N}&  \lambda_{N^*}
\left ( {{m_\pi ^2 f_\pi}\over{m_0}} \right ) \,
\frac{M_N + M_N ^*}{2 M^6} \,
e^{ - \frac{M_N ^2}{M^2}} \,
e^{ - \frac{{M_N ^*}^2}{M^2}} \,
\left ( \Delta g_{\pi N {N^*} } (q^2) \right ) \, =
\nonumber\\
&g&
\left ( \frac{q^2}{M^2} \right )
\frac{\langle \bar{q} q \rangle }{2 \pi^2}
\label{eq23}
\end{eqnarray}
where the function $g$ is defined as:
\begin{equation}
g(x) =  \int_0^1 d\rho
\left [ \frac{x}{(1+\rho)^3} \: + \: 2 \,
\frac{1 + \rho^2 - 4 \rho}{(1 +\rho)^4}
\right ]
e^{ -x \frac{\rho}{1-\rho}}
\label{eq24}
\end{equation}

In order to eliminate the parameters of the higher resonance contribution
as far as possible we can take the derivative
$\frac{\partial}{\partial (\frac{1}{M^2})}$ on both sides of eq.(\ref{eq24})
and substitute the so obtained sum rule back into eq.(\ref{eq24}).
After employing the Gell-Mann--Oakes--Renner relation
\begin{equation}
m_\pi ^2 f_\pi ^2 = - 2 m_0 \langle \bar{q} q \rangle
\label{eq25}
\end{equation}
we obtain finally
\begin{equation}
\Delta g_{\pi NN} (q^2) =
(-) \frac{f_\pi}{M_N}
\frac{M_N ^6}{4 \pi ^2 \lambda_N ^2}
B \left ( q^2,M^2,{M_N ^*}^2 \right )
\label{eq26}
\end{equation}
with
\begin{equation}
B \left ( q^2,M^2,{M_N ^*}^2 \right ) =
\left [
g \left ( \frac{q^2}{M^2} \right )
\frac{{M_N ^*}^2 + M_N ^2 - 3 M^2}
{{M_N ^*}^2 - M_N ^2}
+
\frac{q^2}{{M_N ^*}^2 - M_N ^2}
g^\prime \left ( \frac{q^2}{M^2} \right )
\right ]  \,
\left ( \frac{M}{M_N} \right ) ^6 \,
e^{ 2 \frac{M_N ^2}{M^2}}
\label{eq27}
\end{equation}

Before performing the Borel analysis, let us comment about the region
of $q^2$, where our method can be assumed to work.
If $q^2$ is too low ($q^2 \lesssim \Lambda_{QCD} ^2$)
the OPE breaks down due to higher order power corrections in
$\frac{1}{q^2}$.
On the other hand we have used the PCAC relation (\ref{eq7}) in order to
relate $g_{\pi NN} (q^2)$ to the pseudoscalar nucleon form factor
$g_P (q^2)$ (eq.(\ref{eq8})).
This is equivalent to assume pion pole dominance for $g_P (q^2)$,
i.e. saturating
$\left \langle N(p_1) \vert J_5 ^0 (0) \vert N(p_2) \right \rangle$
with a pion state and using the coupling $g_{\pi NN} (q^2)$.
This assumption likely breaks down if $q^2$ is of
the same magnitude as the first pionic resonance $\pi^\prime$, i.e.
$q^2 \gtrsim m_{\pi^\prime} ^2 \approx 2.2 M_N ^2$, because then
$\pi^\prime$ contributes
in the dispersion relation to
a similar extent than $\pi$ does.

Keeping this in mind we use for the the Borel analysis
$q^2$ values in the interval $0.5 M_N ^2 \lesssim q^2 \lesssim 1.5 M_N ^2$,
where our approach can be assumed to be reliable.
Hereby we follow the spirit of ref.\cite{Hat}
and treat the resonance mass $M_N ^*$ as an effective mass,
which value is adjusted in order to obtain maximal Borel stability.

{}From Figs.(\ref{fig3},\ref{fig4},\ref{fig5}) we can deduce that for the
whole $q^2$ interval mentioned above an effective resonance mass of
${M_N ^*}^2 \approx 6.0 M_N ^2$ ($M_N ^* \approx 2.3 {\mbox{MeV}}$)
gives the largest Borel plateau of
$0.8 M_N ^2 \lesssim M^2 \lesssim 1.8 M_N ^2$.

\section{Results and Discussion}
\label{sec5}

{}From eqs.(\ref{eq26},\ref{eq27}) we obtain the final expression for the
form factor $f_{\pi NN} (q^2)$ defined in eq.(\ref{eq1}):
\begin{equation}
f_{\pi NN} (q^2) = 1 - \frac{f_\pi}{M_N}
\frac{M_N ^6}{g_{\pi NN} 4 \pi ^2 \lambda_N ^2}
B(q^2, M^2 , {M_N ^*}^2)
\label{eq28}
\end{equation}
Due to the discussion in the last section, we use $M^2 = M_N ^2$ and
${M_N ^*}^2 = 6.0 M_N ^2$.
For $M_N$ we take the experimental value $M_N = 940 {\mbox{MeV}}$.
In order to avoid uncertainties from the sum rule for $g_{\pi NN}$
we will also use the experimental value
$g_{\pi NN} = 13.4$ instead of the value from the sum rule
of refs. \cite{RRY1,RRY2,Rei}.
The parameter $\lambda_N$ is not experimentally known but has been
determined from various analyses using the nucleon sum rule,
the most recent and concerning uncertainties obviously most reliable one
coming from ref. \cite{HP}:
\begin{equation}
{\lambda_N}^2  = 5.5 \cdot 10^{-4} {\mbox{GeV}}^6 = 8.07 \cdot 10^{-4} M_N ^6
\label{eq29}
\end{equation}
As we have discussed in the last section formula (\ref{eq29})
cannot be applied for low $q^2$.
This means that we have to interpolate from $q^2 \approx 0.5 M_N ^2$
to $q^2 = 0$, where $f_{\pi NN} (0) =1$.
It is interesting to not that the values for $f_{\pi NN} (q^2)$
which one obtains by this procedure at low $q^2$ are practically
identical to the ones one would get if one applied eq.(\ref{eq29})
literally to the low $q^2$ region.
This might be purely accidental, but a possible reason could be that
higher power corrections in$\frac{1}{q^2}$ blowing up the OPE
at $q^2 \to 0$ are canceled or suppressed after applying the
double Borel transform ${\cal B}_{12}$ (\ref{eq21}).
In order to answer this question one has to examine higher order
power corrections to the OPE diagrams in Fig.\ref{fig2},
which is rather cumbersome for a 3 point function consisting of four
quark lines and will be postponed to a separate analysis.

Anyhow, from Fig.\ref{fig6} we can see that the interpolation
of $f_{\pi NN} (q^2)$ from intermediate $q^2$, where we can apply
eq.(\ref{eq29}), to small $q^2$ is very smooth.
Furthermore in the whole interval $0 < q^2 < 2.0 M_N ^2$ the
$f_{\pi NN} (q^2)$ of our calculation can be fitted very accurately by a
monopole form (\ref{eq1}) with a cutoff mass of
$\Lambda_{\pi} = 0.85 M_N = 800 {\mbox{MeV}}$.
This result is very close to the one recently obtained from
quenched lattice QCD \cite{Liu}, which indicates that both methods
give a similar description of the $\pi NN$ vertex at intermediate $q^2$.

\acknowledgements
This work has been supported by the NSF grant \# PHYS-9310124.
I would like to thank E.Henley (University of Washington)
for various valuable discussions and comments
and the Institute for Nuclear Theory at the University of Washington
for its hospitality during the programs INT-95-1 and INT-95-2.

\begin{figure}
\caption{The three point function $A(p_1,p_2,q)$.}
\label{fig1}
\end{figure}

\begin{figure}
\caption{Diagrams in the OPE}
\label{fig2}
\end{figure}

\begin{figure}
\caption{The Borel curve $B(M^2,q^2,{M_N ^*}^2)$ at $q^2 = 0.5 {M_N}^2$ for
3 values of the resonance mass ${M_N ^*}^2$. Maximum stability is obtained for
${M_N ^*}^2 = 6.0 {M_N}^2$.}
\label{fig3}
\end{figure}

\begin{figure}
\caption{As in (\protect\ref{fig3}) for $q^2 = 1.0 {M_N}^2$.}
\label{fig4}
\end{figure}

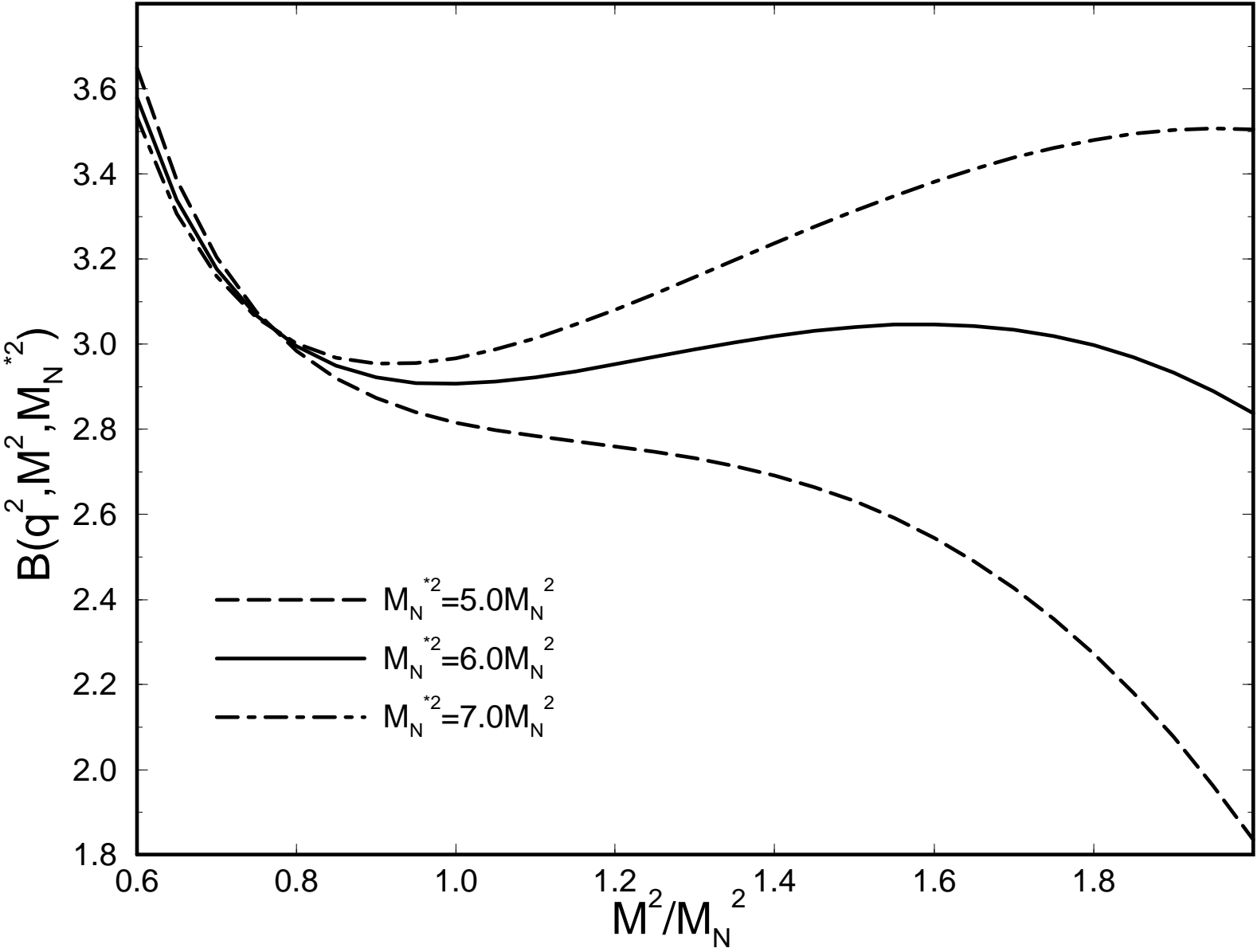
\begin{figure}
\caption{As in (\protect\ref{fig2}) for $q^2 = 1.5 {M_N}^2$.}
\label{fig5}
\end{figure}

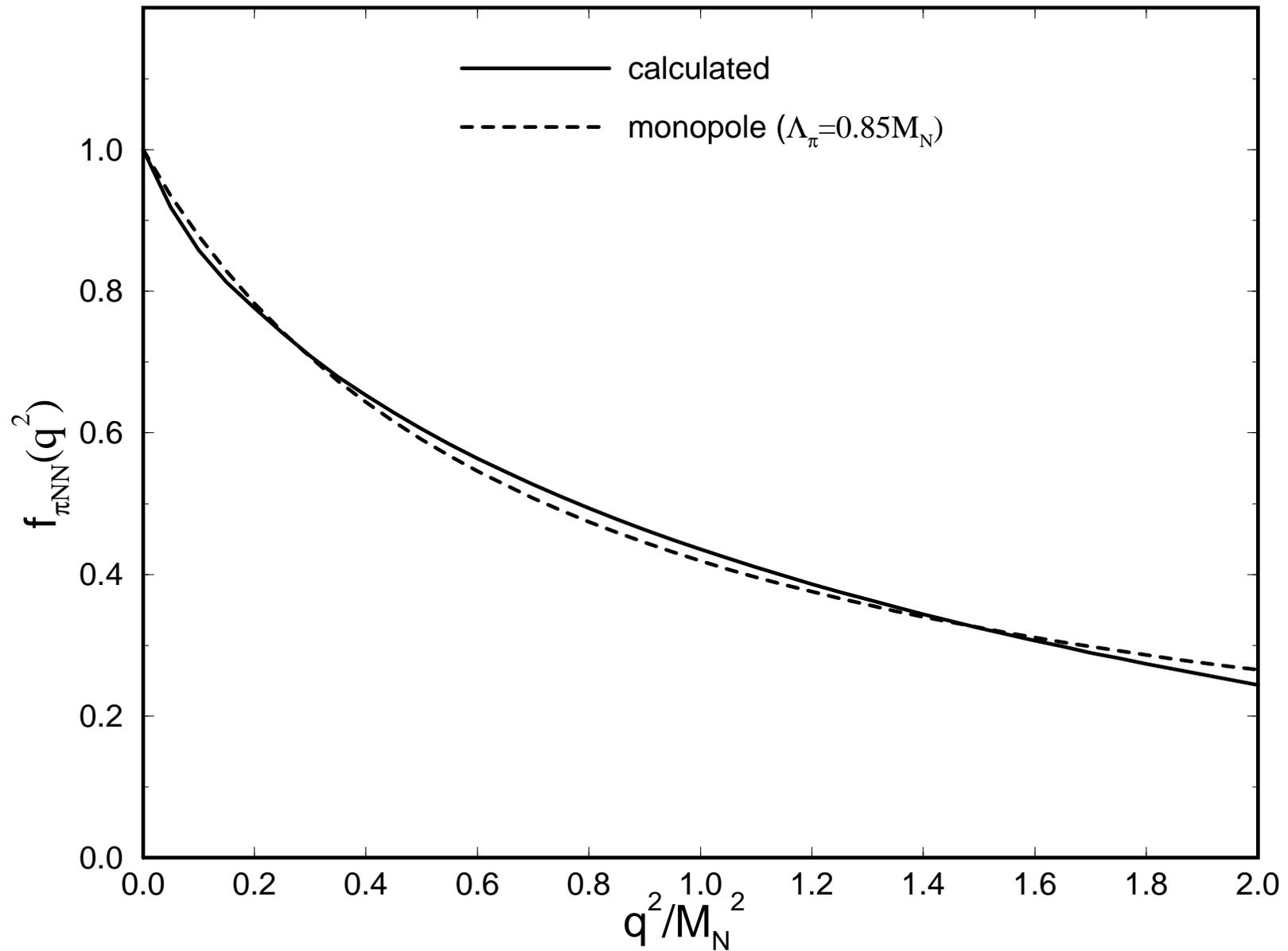
\begin{figure}
\caption{The $f_{\pi NN} (q^2) $ calculated from
eq.(\protect{\ref{eq29}}) (interpolated to small $q^2$) with
$M^2 = {M_N}^2$ and ${M_N ^*}^2 = 6.0 {M_N}^2$
(full line) compared with a monopole form
(\protect{\ref{eq1}}) with a cutoff mass of
$\Lambda_{\pi} = 0.85 M_N = 800 \ {\mbox{MeV}}$ (dotted line).}
\label{fig6}
\end{figure}

\end{document}